\documentclass[12pt]{spieman}  
\usepackage{amsmath,amsfonts,amssymb}
\usepackage{multirow, multicol, tabularx}
\usepackage{graphicx}
\usepackage{setspace}
\usepackage{tocloft}
\usepackage{float}
\usepackage{hyperref}
\title{SELMA3D challenge: Self-supervised learning for 3D light-sheet microscopy image segmentation}

\author[a,b*]{Ying Chen}
\author[a, c]{Rami Al-Maskari}
\author[a, c]{Izabela Horvath}
\author[a]{Mayar Ali}
\author[a]{Luciano Höher}
\author[f]{Kaiyuan Yang}
\author[g]{Zengming Lin}
\author[h]{Zhiwei Zhai}
\author[h]{Mengzhe Shen}
\author[i]{Dejin Xun}
\author[i]{Yi Wang}
\author[j]{Tony Xu}
\author[j]{Maged Goubran}
\author[k]{Yunheng Wu}
\author[k,l,m]{Kensaku Mori}
\author[d,e*]{Johannes C. Paetzold}
\author[a,b*]{Ali Erturk}
\affil[a]{Institute for Tissue Engineering and Regenerative Medicine, Helmholtz Center Munich, German Research Center for Environmental Health, Neuherberg, Germany}
\affil[b]{Institute for Stroke and Dementia Research, Klinikum der Universität München, Ludwig-Maximilians University Munich, Munich, Germany}
\affil[c]{TUM School of Computation, Information and Technology (CIT), Technical University of Munich, Munich, Germany}
\affil[d]{Department of Computing, Imperial College London, United Kingdom}
\affil[e]{Department of Radiology, Weill Cornell Medicine, Cornell University, New York, United States}
\affil[f]{Department of Quantitative Biomedicine, University of Zurich, Zurich, Switzerland}
\affil[g]{Shanghai University of Finance and Economics, Shanghai, China}
\affil[h]{BGI Research, Shenzhen, China}
\affil[i]{National Key Laboratory of Chinese Medicine Modernization, Innovation Center of Yangtze River Delta, Zhejiang University, Jiaxing, China}
\affil[j]{Sunnybrook Research Institute, University of Toronto, Toronto, Canada}
\affil[k]{Graduate School of Informatics, Nagoya University, Nagoya, Japan}
\affil[l]{Information Technology Center, Nagoya University, Nagoya, Japan}
\affil[m]{Research Center for Medical Bigdata, National Institute of Informatics, Tokyo, Japan}

\cftpagenumbersoff{figure}
\cftpagenumbersoff{table} 
\begin{document} 
\maketitle

\begin{abstract}
Recent innovations in light sheet microscopy, paired with developments in tissue clearing techniques, enable the 3D imaging of large mammalian tissues with cellular resolution. Combined with the progress in large-scale data analysis, driven by deep learning, these innovations empower researchers to rapidly investigate the morphological and functional properties of diverse biological samples. Segmentation, a crucial preliminary step in the analysis process, can be automated using domain-specific deep learning models with expert-level performance. However, these models exhibit high sensitivity to domain shifts, leading to a significant drop in accuracy when applied to data outside their training distribution. To address this limitation, and inspired by the recent success of self-supervised learning in training generalizable models, we organized the SELMA3D Challenge during the MICCAI 2024 conference\footnote{https://selma3d.grand-challenge.org}. SELMA3D provides a vast collection of light-sheet images from cleared mice and human brains, comprising 35 large 3D images-each with over $1000^3$ voxels-and 315 annotated small patches for finetuning, preliminary testing and final testing. The dataset encompasses diverse biological structures, including vessel-like and spot-like structures. Five teams participated in all phases of the challenge, and their proposed methods are reviewed in this paper. Quantitative and qualitative results from most  participating teams demonstrate that self-supervised learning on large datasets improves segmentation model performance and generalization. We will continue to support and extend SELMA3D as an inaugural MICCAI challenge focused on self-supervised learning for 3D microscopy image segmentation.
\end{abstract}


{\noindent \footnotesize\textbf{*}Ying Chen, \linkable{Ying.Chen@campus.lmu.de}; Johannes C. Paetzold,  \linkable{jpaetzold@med.cornell.edu}; \\ Ali Erturk, \linkable{ali.erturk@helmholtz-munich.de}}

\begin{spacing}{2}   

\section{Introduction}
\label{sect:intro}  
In the realm of modern biological research, the ability to visualize and understand complex structures within tissues and organisms is crucial. Traditional imaging methods often face challenges in providing a cellular resolution [\citenum{Yoshii17}], 3D view [\citenum{Jahn20}] of bio-samples while preserving their integrity [\citenum{Maes22}]. The integration of tissue clearing and light-sheet microscopy (LSM) overcomes these limitations, serving as an efficient, high contrast, and ultra-high resolution method for visualizing a wide range of biological structures in diverse samples, such as cellular and subcellular structures, organelles and processes [\citenum{Stelzer21}].

Tissue clearing techniques render inherently opaque biological samples transparent, allowing light to penetrate deeper into the tissue [\citenum{Ueda21}] and imaging reagents (e.g., fluorophores or antibodies), while preserving their structural integrity and molecular content. Various fluorophores or antibodies can be employed to selectively stain specific biological structures within samples and enhance their contrast under microscopy [\citenum{Poola19}]. After staining and tissue clearing, LSM provides rapid 3D imaging of intricate biological structures with high spatial resolution, offering new insights into various biomedical research fields such as neuroscience [\citenum{Ueda20}], immunology [\citenum{zhang24}], oncology [\citenum{Almagro21}] and cardiology [\citenum{Fei16}].

With automated image analysis approaches, scientists can extract structural and functional cellular and subcellular information from LSM images of various bio-samples at an accelerated pace [\citenum{Ueda21, Tian21}]. To analyze LSM images, segmentation plays a pivotal and essential role in identifying and distinguishing various biological structures in different biomedical research fields [\citenum{Amat15}]. For whole-organ or body LSM images, manual segmentation is time-intensive, single images can have $10000^3$ voxels, hence automatic segmentation methods are required. Recent strides in deep learning-based segmentation models offer promising solutions for automated segmentation of LSM images [\citenum{Kumar23, Todorov20,Shit21}]. Although these models reached segmentation performances comparable to expert human annotators, their success largely relies on supervised learning from substantial manual annotations of high quality, and these models are usually task-specific with limited generalizability to different applications [\citenum{Zhou23}]. Therefore, the widespread use of deep learning-based segmentation models is constrained as the annotation for every specific LSM image segmentation task requires experts with domain knowledge, making it impractical. 

Hence, we propose that it is crucial to develop generalizable models capable of serving multiple LSM image segmentation tasks. Recent years have witnessed remarkable success towards developing foundation models (FMs) in various domains, from natural language processing [\citenum{Touvron23}], computer vision [\citenum{Wang23}], audio and speech processing [\citenum{Awais23}], to biology and healthcare [\citenum{Krishnan22, Zhang24b, Zhao24}]. FMs are large deep learning neural networks trained on extensive data sets, exhibiting considerable generalizability and adaptability. Self-supervised learning, a paradigm where the model learns general representations from unannotated data through ‘pretext tasks’ for which labels are not required or can be generated automatically [\citenum{Chen20, He22, Oquab23}], proves advantageous in the pre-training of FMs with massive unannotated data. Subsequently, the model can be fine-tuned for specific tasks [\citenum{Krishnan22}].

Notably, self-supervised learning has not been extensively explored within the LSM field, despite the presence of vast sets of LSM data of different biological structures. Some of the properties of LSM images, e.g. the high signal-to-noise ratio, make the data specifically well suited for self-supervised learning. To fill this gap, we host an inaugural MICCAI challenge on self-supervised learning for 3D LSM image segmentation (SELMA3D), encouraging the community to develop self-supervised learning methods for general segmentation of various structures in 3D LSM images. With an effective self-supervised learning method, extensive 3D LSM images with no annotations can be leveraged to pretrain segmentation models. This encourages models to capture high-level representations that are generalizable across different biological structures. Subsequently, the pretrained models can be finetuned on substantially smaller annotated datasets, thereby significantly minimizing the annotation efforts in various 3D LSM segmentation applications.

The contribution of our organized SELMA3D challenge can be summarized below:
\begin{itemize}
    \item SELMA3D marks a significant milestone in advancing research on self-supervised learning in the domain of 3D microscopy images. To the best of our knowledge, it is the first challenge to offer a large-scale dataset encompassing images of diverse biological structures and specimens for model pretraining, addressing a critical gap in this field.
    \item SELMA3D focuses on segmentation as the downstream task for assesing pretrained models developed through self-supervised learning, as segmentation plays a critical role in most analysis workflows. To support this, we further design an annoated dataset for model fine-tuning and evaluation, covering both vessel-like and spot-like structures.
    \item SELMA3D pays attention to the generalizability of the developed models. By leveraging self-supervised learning for pretraining, the segmentation models are expected to generalize to unseen data. To assess this capability, we introduce a new biological structure during the test phase. 
\end{itemize}

In the paper, we will first summarize previous work related to light-sheet microscopy image segmentation and self-supervised learning strategies. Then we describe our SELMA3D challenge dataset and present the dataset properties. We then give an overview of the SELMA3D challenge setup and evaluation methods. We summarize the submitted methods and their performance on our test set. Lastly, we discuss algorithm designs, open problems, and future tasks.

\section{Related Works}
\subsection{Light-sheet Microscopy Smage Segmentation}
Light-sheet microscopy (LSM) can achieve imaging at cellular and subcellular resolution, making it a widely utilized technology across numerous areas of life sciences, e.g. observing embryo evolution and development [\citenum{Blain23}] , mapping neuronal networks [\citenum{Kaltenecker24}], investigating cancer metastasis [\citenum{Pan19}]. For analyzing features of biological structures in these LSM images, segmentation is a crucial step. Compared to manual segmentation [\citenum{Wang19}] or traditional image segmentation approaches [\citenum{Packard17, Spangenberg23}], deep learning-based segmentation methods offer greater efficiency and accuracy, continuously gaining popularity in different LSM image segmentation tasks [\citenum{Pan19, Todorov20, Friedmann20, Kaltenecker24}]. However, most deep learning methods are constrained to one specific biological structure, limiting their broader applicability. Besides, the lack of publicly available annotated LSM image datasets requires researchers to generate their annotations for model training, posing a significant barrier. These challenges hinder the widespread usage of deep learning for advancing LSM image analysis.   

\subsection{Self-supervised Learning for Images}
Self-supervised learning 
(SSL) leverages unlabeled data to learn underlying and robust feature representation through the design of pretext tasks. These pretext tasks enable harnessing the power of unlabeled data and enhance the model's generalizability and performance [\citenum{Krishnan22, Zhou23, Kirillov23}]. Pretext tasks in SSL methods for image domain can generally be categorized into three groups: context-based methods, generative methods, contrastive methods [\citenum{Gui24}]. Common context-based pretext tasks include rotation prediction, jigsaw puzzle solving, patch location [\citenum{Taleb20}] and so on. Contrastive pretext tasks, i.e. contrastive learning, is a cutting-edge branch of SSL. They aim to learn an embedding space where similar samples (positive pairs) are close to each other while dissimilar ones (negative pairs) are pushed apart [\citenum{Chen20, Oquab23}]. In recent years, generative pretext tasks have gained significant popularity thanks to the emergence of advanced masked image modeling (MIM) methods such as masked auto-encoder (MAE) [\citenum{He22}] and BEiT [\citenum{Tian23}], posing a challenge to the dominance of contrastive learning. The core idea behind MIM methods is to extract visual representations from unlabeled image data by predicting missing pixels from masked portions of images.  

SSL has been successfully applied in the development of foundation models across a wide range of domains, including natural imagery [\citenum{Kirillov23, Wang23}], medical imaging [\citenum{Zhou23, Zhao24}], digital pathology [\citenum{Chen24, Xu24}], and beyond. Despite several attempts to apply self-supervised learning (SSL) to microscopy images [\citenum{Achard2024, Zheng24}], its use in developing foundation models with large-scale 3D microscopy image data remains less explored. 

\begin{table}[]
\caption{Details of the training set without ground truth anottations.} 
\label{tab:ublabeled}
\begin{center}       
\begin{tabular}{l|l|c|c} 
\hline
\rule[-1.5ex]{0pt}{5ex} \textbf{Biological structure} & \textbf{Specimen source} & \textbf{Resolution$(X\times Y \times Z)$} & \textbf{Size$(X\times Y \times Z)$}  \\
\hline\hline
\multirow{9}{*}{Blood vessel} & \multirow{9}{*}{mouse brain} & \multirow{9}{*}{$1.625um\times1.625um\times3um$} & {\small $4786\times3108\times2073$} \\
\rule[-1ex]{0pt}{3ex}       &                          &                          & {\small $4783\times3109\times1981$} \\
\rule[-1ex]{0pt}{3ex}       &                          &                          & {\small $4772\times3109\times1870$} \\
\rule[-1ex]{0pt}{3ex}       &                          &                          & {\small $3661\times2804\times2012$} \\
\rule[-1ex]{0pt}{3ex}       &                          &                          & {\small $3837\times2954\times2083$} \\
\rule[-1ex]{0pt}{3ex}       &                          &                          & {\small $4120\times3138\times2086$} \\
\rule[-1ex]{0pt}{3ex}       &                          &                          & {\small $4948\times3227\times1915$} \\
\rule[-1ex]{0pt}{3ex}       &                          &                          & {\small $3823\times3196\times2023$} \\
\rule[-1ex]{0pt}{3ex}       &                          &                          & {\small $3935\times3336\times1943$} \\
\hline
\multirow{18}{*}{c-Fos$+$ cells} & \multirow{18}{*}{mouse brain} & \multirow{18}{*}{$1.625um\times1.625um\times6um$} & {\small $6656\times5616\times733$} \\
\rule[-1ex]{0pt}{3ex}       &                          &                          & {\small $6656\times5616\times715$} \\
\rule[-1ex]{0pt}{3ex}       &                          &                          & {\small $6656\times5616\times737$} \\
\rule[-1ex]{0pt}{3ex}       &                          &                          & {\small $6656\times5616\times681$} \\
\rule[-1ex]{0pt}{3ex}       &                          &                          & {\small $6656\times5616\times700$} \\
\rule[-1ex]{0pt}{3ex}       &                          &                          & {\small $6656\times5616\times669$} \\
\rule[-1ex]{0pt}{3ex}       &                          &                          & {\small $6656\times5616\times726$} \\
\rule[-1ex]{0pt}{3ex}       &                          &                          & {\small $6656\times5616\times679$} \\
\rule[-1ex]{0pt}{3ex}       &                          &                          & {\small $6656\times5616\times715$} \\
\rule[-1ex]{0pt}{3ex}       &                          &                          & {\small $6656\times5616\times733$} \\
\rule[-1ex]{0pt}{3ex}       &                          &                          & {\small $6656\times5616\times721$} \\
\rule[-1ex]{0pt}{3ex}       &                          &                          & {\small $6656\times5616\times693$} \\
\rule[-1ex]{0pt}{3ex}       &                          &                          & {\small $6656\times5616\times712$} \\
\rule[-1ex]{0pt}{3ex}       &                          &                          & {\small $6656\times5616\times713$} \\
\rule[-1ex]{0pt}{3ex}       &                          &                          & {\small $6656\times5616\times714$} \\
\rule[-1ex]{0pt}{3ex}       &                          &                          & {\small $6656\times5616\times734$} \\
\rule[-1ex]{0pt}{3ex}       &                          &                          & {\small $6656\times5616\times677$} \\
\rule[-1ex]{0pt}{3ex}       &                          &                          & {\small $6656\times5616\times711$} \\
\hline
\multirow{4}{*}{cell nucleus} & \multirow{4}{*}{human brain} & \multirow{4}{*}{$0.54um\times0.54um\times5um$} & {\small $19655\times11855\times834$} \\
\rule[-1ex]{0pt}{3ex}       &                          &                          & {\small $13478\times8133\times613$} \\
\rule[-1ex]{0pt}{3ex}       &                          &                          & {\small $13307\times17238\times611$} \\
\rule[-1ex]{0pt}{3ex}       &                          &                          & {\small $16054\times16565\times768$} \\
\hline
\multirow{4}{*}{A$\beta$ plaques} & \multirow{4}{*}{mouse brain} & \multirow{4}{*}{$1.63um\times1.63um\times4um$} & {\small $4850\times4237\times2166$} \\
\rule[-1ex]{0pt}{3ex}       &                          &                          & {\small $5038\times4671\times2230$} \\
\rule[-1ex]{0pt}{3ex}       &                          &                          & {\small $4990\times4258\times2360$} \\
\rule[-1ex]{0pt}{3ex}       &                          &                          & {\small $4692\times3456\times2268$} \\
\hline
\end{tabular}
\end{center}
\end{table} 

\begin{table}[h]
\caption{Details of patches with ground truth anottations in the training, preliminary test and final test sets} 
\label{tab:labeled}
\begin{center}       
\begin{tabular}{l l c c} 
\hline
\rule[-1.5ex]{0pt}{5ex} \textbf{Data split} & \textbf{Biological structure} & \textbf{Patch Size$(X\times Y \times Z)$}  & \textbf{Patch number} \\
\hline
\hline
\multirow{4}{*}{Training set} & {Blood vessel} & {$500\times500\times50$} & {24} \\
\rule[-1ex]{0pt}{3ex}       & {c-Fos$+$ cells} & {$100\times100\times100$} & {19} \\
\rule[-1ex]{0pt}{3ex}       & {cell nucleus} & {$200\times200\times200$} & {12} \\
\rule[-1ex]{0pt}{3ex}       & {A$\beta$ plaques} & {$300\times300\times300$} & {34} \\
\hline
\multirow{2}{*}{Preliminary test set} & {c-Fos$+$ cells} & {$100\times100\times100$} & {23} \\
\rule[-1ex]{0pt}{3ex}       & {Microglia} & {$100\times100\times100$} & {85} \\
\hline
\multirow{2}{*}{Final test set} & {c-Fos$+$ cells} & {$200\times200\times200$} & {8} \\
\rule[-1ex]{0pt}{3ex}       & {Microglia} & {$200\times200\times200$} & {8} \\
\hline
\end{tabular}
\end{center}
\end{table}

\section{Challenge Setup}
\subsection{Challenge Data Cohort}
The challenge dataset consisted of mouse and human brain images, collected by the Institute for Tissue Engineering and Regenerative Medicine (iTERM) and the Institute for Stroke and Dementia Research between 2019 and 2023. The dataset encompasses diverse stained biological structures, including blood vessels [\citenum{Todorov20}], c-Fos $+$ cells [\citenum{Kaltenecker24}], cell nuclei [\citenum{Zhao20}] and amyloid-beta (A$\beta$) plaques [\citenum{Bhatia22}]. Based on morphology, these structures are primarily categorized into two types: tree-like tubular structures, e.g. vessel, and spot-like structures, e.g. c-Fos$+$ cells, cell nuclei, and amyloid-beta (A$\beta$) plaques.    

\subsection{Image Data Acquisition}
The data acquisition process followed a standardized workflow: structure staining, tissue clearing, and LSM imaging. Various stains were used to selectively bind to specific structures within the samples, improving their contrast against the surrounding tissue. Four stains are involved in the image acquisition process for this study: wheat germ agglutinin (WGA) and Evans blue (EB) for visualizing blood vessels, c-Fos staining for cells involved in neuronal activity, TO-PRO-3 staining for cell nuclei, Congo Red staining for amyloid-beta (A$\beta$) plaques. Two tissue clearing methods were adopted as described in our prior works [\citenum{Zhao20, Erturk12}]. For further details on staining, tissue clearing, and other sample preparation steps, please refer to the original work [\citenum{Todorov20, Kaltenecker24, Zhao20, Bhatia22}].

After sample preparation, LSM imaging was performed using an UltraMicroscope II (LaVision BioTec) or prototype UltraMicroscope (Miltenyi Biotec) coupled to a white light laser module (NKT SuperK Extreme EXW-12).


\subsection{Data Annotation Setup}
The manual annotation and verification processes are conducted in 3D using virtual reality (VR) for visualization efficiency. Each case undergoes a hierarchical annotation process, beginning with initial semantic segmentation annotations performed manually by 4 expert annotators experienced in LSM imaging. The initial manual annotations are conducted by expert annotators with in-depth biological and anatomy training, ensuring a comprehensive understanding of LSM. Subsequently, an expert with three years of professional experience in LSM reviews and refines the initial annotations. The final annotations are then determined and approved by two senior experts with five or more years of professional experience in LSM.

\subsection{Dataset Split}
The whole dataset comprises a training set, preliminary test set, and final test set. The training set has two portions. The first portion includes a large set of 3D LSM images of both mouse and human brains without annotations, intended for model pretraining through self-supervised learning. This portion consists of 9 3D images of blood vessels from whole mouse brains, 18 of c-Fos $+$ cells from whole mouse brains, 4 of amyloid-beta (A$\beta$) plaques from whole mouse brains, 4 of cell nuclei from 4 human brain sub-regions (Hippocampus, motor cortex, sensory cortex, and visual cortex). Details of every individual 3D image are listed in Table \ref{tab:ublabeled}. The second portion contains cropped patches of brain 3D LSM images representing the four biological structures mentioned above, accompanied by precise annotations. These annotated patches enable the fine-tuning of the model for semantic segmentation tasks. 

To assess the model's generalization during the preliminary test and final test phases, evaluations will be performed using 3D brain LSM patches representing two types of biological structures. Besides, while half of the patches contain biological structures already shown in the training set, the other half introduces a new biological structure, that is, microglia cells. Microglia cells are characterized by branch-like extensions resembling vessels and a small cellular body. Finally, the preliminary test set includes 23 c-Fos$^+$ cell patches and 85 microglia patches of size $100\times100\times100$, while the test set contains 8 c-Fos$^+$ cell patches and 8 microglia patches of size $200\times200\times200$. A summary of all annotated patches in the training, preliminary test, and final test sets is presented in Table \ref{tab:labeled}.

For the unannotated portion of the training set, given the large size of every brain LSM image, each 2D slice of the large 3D brain LSM images is saved as a 16-bit signed TIFF file. For the annotated portion of the training set, as well as the preliminary test and final test sets, all small patches along with their corresponding annotations are stored in NIfTI format with 16-bit signed precision and in LPS+ orientation.

\subsection{Evaluation Methods}
The biological structures in this challenge fall into two categories: spot-like structures and tree-like tubular structures. To evaluate the generalization capability of segmentation models, the results will be assessed separately for each structure type during the preliminary test and final test phases. To be specific, for spot-like structures, i.e. c-Fos$^+$ cells, segmentation results are evaluated by 2 metrics: volumetric Dice similarity coefficient and Betti number error in dimension 0. For tree-like tubular structures, i.e. microglia, 4 metrics are utilized: volumetric Dice similarity coefficient, Betti number error in dimension 0, Betti  number error in dimension 1, and centerline Dice similarity coefficient.

For spot-like structure segmentation task, the objective is to accurately detect individual signals. The segmentation of tree-like structures focuses on preserving anatomical topology. The Dice similarity coefficient evaluates voxel overlap between the ground truth and segmentation result. The Betti number errors assess the differences in topological features, such as connected components and circular holes, between the ground truth and the segmentation result [\citenum{Stucki23}]. The Centerline-Dice (clDice) metric specifically measures voxel-wise overlap in tubular and curvilinear structures, measuring how well the predicted segmentation captures tree-like structures [\citenum{Shit21}].

\section{Participating Methods}
This section begins with an overall comparison of the methods submitted by different participants. Next, we summarize the key techniques and provide detailed implementation information for each method. The methods are presented in order of their performance ranking in the final test phase. Finally, we discuss potential directions for improvement.
\subsection{General Participation}
SELMA3D attracted 84 registered participants from four continents. 5 teams successfully submitted solutions to both preliminary and final test phases of the challenge. Table \ref{tab:methods} summarizes the key characteristics of the different methods, including the self-supervised learning strategy, data preprocessing steps for the unannotated dataset, the network architecture used during the self-supervised learning stage, the fine-tuning strategy, preprocessing of annotated patches, and the final fine-tuned segmentation network. While different segmentation networks are utilized by different teams, contrastive learning methods and masked volume inpainting are commonly employed self-supervised learning strategies. Given the large size of the unannotated dataset, 4 out of 5 teams implemented preprocessing steps to reduce training time, either by downsampling the images or selecting parts of images. 

\begin{table}[H]
\caption{A brief summary of different methods, encompassing the training strategies, data preprocessing approaches and networks involved in self-supervised learning and finetuning stages.} 
\label{tab:methods}
\hspace*{-1in}
\begin{tabular}{p{1.3cm}|p{3cm}p{3.5cm}p{1.7cm}||p{3.5cm}p{3.5cm}p{1.7cm}} 
\hline
\rule[-1.5ex]{0pt}{4ex} \multirow{2}{*}{\textbf{Team}} & \multicolumn{3}{c||}{\textbf{Self-supervised learning stage}} & \multicolumn{3}{c}{\textbf{Finetuning stage}} \\
\rule[-1.5ex]{0pt}{3ex} & {\footnotesize SSL strategy} & {\footnotesize data preprocessing} & {\footnotesize network} & {\footnotesize finetune strategy} & {\footnotesize data preprocessing} & {\footnotesize network}  \\
\hline
\rule[-1.5ex]{0pt}{4ex} {\footnotesize Zoomlin} & {\footnotesize BYOL} & {\footnotesize Converted 2D slice images from TIFF to JPG format; Stack 2D images to form 3D volumes} & {\footnotesize 3D UNet encoder} & {\footnotesize While the encoder was frozen, the decoder was finetuned using Dice loss} & {\footnotesize None} & {\footnotesize 3D UNet}\\
\hline
\rule[-1.5ex]{0pt}{4ex} {\footnotesize BioAI} & {\footnotesize masked volume inpainting, image rotation prediction, contrastive predictive coding (CPC) and biological structure classification} & {\footnotesize 2k 3D sub-volumes of $400^3$ were randomly selected and cropped from the large LSM images. 3D patches of size $96^3$ were further extracted from these sub-volumes to train models} & {\footnotesize SwinUNETR encoder} & {\footnotesize The whole SwinUNETR was finetuned using a combination of focal loss, Dice loss and boundary loss.} & {\footnotesize None} & {\footnotesize SwinUNETR}\\
\hline
\rule[-1.5ex]{0pt}{4ex} {\footnotesize XunDJ} & {\footnotesize masked volume inpainting} & {\footnotesize 2D plane images were downsampled to $1/3$ of the original width and height. After normalization, 3D patches were randomly cropped from the large 3D image of size $64^3$} & {\footnotesize SwinUNETR} & {\footnotesize The model was fintuned using Dice loss.} & {\footnotesize Data augmentation including  normalization, random cropping, padding, random flipping, random rotation, and random intensity variation} & {\footnotesize SwinUNETR}\\
\hline

\rule[-1.5ex]{0pt}{4ex} {\footnotesize Tonyxu} & {\footnotesize DINOV2 + iBOT, extended to 3D} & {\footnotesize 3D patches of size $128^3$ were cropped from the unannotated dataset. They applied a foreground intensity threshold to create a foreground mask, then filtered out patches if less than $10\%$ of its voxels were foreground. Consequently, approximately 300k patches were obtained to pretrain the model.} & {\footnotesize 3D ViT} & {\footnotesize With the ViT weights kept frozen, the adapter and decoder were fine-tuned using a combination of Dice loss and cross-entropy loss} & {\footnotesize Patch intensities were scaled from the $0.05^{th}$ and $99.95^{th}$ percentiles to the range $[-1, 1]$. Random augmentations were applied, including random affine transforms, random contrast adjustment, random additive Gaussian noise, and random flipping} & {\footnotesize 3D ViT + adapter + convolutional decoder}\\
\hline
\rule[-1.5ex]{0pt}{4ex} {\footnotesize Wu} & {\footnotesize SimCLR} & {\footnotesize The original 2D image planes were downsampled to 1/4 of their size.} & {\footnotesize Attention U-Net encoder}  & {\footnotesize A combination of focal loss and Dice loss was utilized to finetuned the whole network.} & {\footnotesize 3D patches were converted into 2D slices and every slice was resized to $256^2$ before feeding into the network. } & {\footnotesize Attention U-Net}\\
\hline
\end{tabular}

\end{table}

\subsection{Participants Methods}
\subsubsection{Participating Team - \textbf{A. Zoomlin}}
The Zoomlin team adopted BYOL (Bootstrap Your Own Latent) contrastive learning [\citenum{Grill20}] for self-supervised training on unlabeled 3D microscopic images. This contrastive learning framework employs two neural networks: the online network and the target network. The online network is comprised of three components: an encoder $f_\theta$, a projector $g_\theta$, and a predictor $q_\theta$. The target network mirrors the architecture of the online network but lacks the predictor and operates with a distinct set of weights $\xi$, ensuring stability and avoiding collapse. The target network is updated via an exponential moving average (EMA) of the online network’s weights. The contrastive learning process involves augmenting input images to create two different views. The two views are separately fed into the online and target networks to extract feature maps. A contrastive loss, specifically the mean squared error between features of the two sets of feature maps, is used to maximize the similarity between augmented views of the same image. 

For the implementation, they utilized 3D UNet [\citenum{Isensee21}] as the backbone architecture. The encoder was pre-trained using self-supervised learning, enabling it to extract meaningful features of various biological structures from the unlabeled data. During the fine-tuning stage, the encoder was frozen while the decoder was trained with labeled data. This supervised fine-tuning step mapped the encoder’s learned features to the correct labels, optimizing the network for 3D light-sheet microscopic image segmentation. In addition, Zoomlin team converted 2D slice images from TIFF to JPG format to reduce file size and computational load while retaining the essential visual information necessary for effective feature learning. Then, 2D images are stacked to form 3D volumes for model learning spatial hierarchies and contextual information. 

\subsubsection{Participating Team - \textbf{bioAI}}
The bioAI team built upon the study of Tang et al.[\citenum{Tang22}] for self-supervised learning, which incorporated multiple pretext tasks, including masked volume inpainting, image rotation prediction, and contrastive predictive coding (CPC). Additionally, they introduced a new pretext task: classifying different biological structures within the images. The backbone architecture used for their approach was SwinUNETR [\citenum{Tang22}]. For pretraining, 90,000 3D sub-volumes of approximately $400\times400\times400$ were cropped from the 3D LSM images. Due to computational constraints, about 2,000 3D sub-volumes were randomly selected for training. During the self-supervised learning stage, 3D patches of size $96\times96\times96$ were further extracted from these sub-volumes, utilized to train models for 100,000 iterations.

During fine-tuning, the bioAI team addressed the class imbalance between the labeled biological structures and the background by combining focal loss [\citenum{Lin17}], dice loss and boundary loss [\citenum{Kervadec21}]. To enhance segmentation performance, they employed additional strategies. Given that the preliminary test and final test phases involved two types of biological structures—c-Fos$+$ cells and microglia—with only c-Fos$+$ cells present in the training set while microglia data were unseen, the team introduced a DenseNet to classify whether the test data belonged to c-Fos$+$ cells or not. For c-Fos$+$ cell segmentation, they fine-tuned a dedicated segmentation model trained solely on c-Fos$+$ cell data. For microglia segmentation, they fine-tuned a separate network using a mixture of all annotated data. Additionally, a post-processing module was implemented to remove small isolated regions with fewer than 5 voxels after obtaining segmentation predictions. For fine-tuning, the annotated dataset was split into training ($80\%$), preliminary test ($10\%$) and final test ($10\%$) subsets. The fine-tuning process was carried out over 500 epochs.
 
\subsubsection{Participating Team - \textbf{XunDJ}}
The XunDJ team introduced masked volume inpainting as the pretext task for self-supervised learning. They utilized SwinUNETR as the backbone architecture. For self-supervised learning, preprocessing steps were applied before training. Initially, 2D plane images were downsampled to one-third of their original width and height and converted to PNG format to optimize memory usage. A sample dictionary was then generated to enable random sampling of continuous sets of plane images along the z-dimension, creating a large 3D volume. Finally, after normalization, 3D patches were randomly cropped from the large 3D image and padded to achieve a uniform size $64\times64\times64$. The pretraining lasted for 100 epochs with a batch size of 4.

For fine-tuning, several augmentation techniques were applied including normalization, random cropping, padding, random flipping, random rotation, and random intensity variation. The Dice loss was used as the loss function. Training was conducted for 2000 epochs with a batch size of 4.

\subsubsection{Participating Team - \textbf{Tonyxu}}
The Tonyxu team adopted the self-supervised learning method DINOV2 [\citenum{Oquab23}] to pretrain an encoder network. DINOv2 leverages an exponential moving average teacher and a self-distillation objective for pretraining. DINOv2 was originally designed for natural images. The Tonyxu team adapted it for 3D inputs, employing the 3D Vision Transformer (ViT) network [\citenum{Dosovitskiy20}]as the encoder. Besides, they employed a patch-level masked image modeling objective adapted from iBOT [\citenum{Zhou21}] to boost pretraining for segmentation. They extract 3D patches of size $128\times128\times128$ from the unannotated dataset. To boost pretraining efficiency, they filtered out background regions in the image. They applied a foreground intensity threshold of 500 (arbitrary units) to create a foreground mask, selecting out patches only if more than $10\%$ of the voxels were foreground. Consequently, they obtained approximately 300,000 patches, which were used to pretrain the model for 125,000 iterations over roughly three days.

During the finetuning stage, a lightweight adapter module [\citenum{Chen22}] and a convolutional decoder were added to the ViT encoder to produce segmentation masks. The fine-tuning dataset was randomly divided into three cross-validation folds, stratified based on the type of biological structure in the images. For intensity normalization, the image intensities were scaled from the $0.05^{th}$ and $99.95^{th}$ percentiles to the range $[-1, 1]$. To enhance the network's generalizability to unseen data, extensive random augmentations were applied, including random affine transforms, random contrast adjustment, random additive gaussian noise, and random flipping. With the ViT weights kept frozen, the adapter and decoder were fine-tuned for 30,000 iterations using a combination of Dice loss and cross-entropy loss with the AdamW optimizer. The learning rate was warmed up to $1\times 10^{-4}$ over the first 3,000 iterations.

\subsubsection{Participating Team - \textbf{Wu}}
Different from other teams, the Wu team developed a segmentation model based on 2D images. They utilized a 2D Attention U-Net as the backbone. During the self-supervised learning stage, they adopted the contrastive learning method SimCLR [\citenum{Chen20}]. Specifically, two different transformations—contrast enhancement and Gaussian blur—were applied to each image slice. The goal of SimCLR is to make the features of the two augmented versions of the same image more similar while encouraging greater dissimilarity between features of different augmented images. To reduce training time, they downsampled the original images to 1/4 of their size. During the fine-tuning stage, they converted 3D patches into 2D slices to feed into the network. To address the class imbalance issue, they combined focal loss and Dice loss as the final loss function for the model's supervised learning.

\setlength{\tabcolsep}{2.5pt}
\setlength{\belowcaptionskip}{6pt}
\begin{table}[t]
\caption{Quantitative results on the prelimitary test sets achieved by participants. The results are reported in the format of mean $\pm$ standard deviation. The arrows indicate better performance for each metric.} 
\label{tab:pre_results}
\centering
\begin{tabular}{c c c | c c c c} 
\hline
\rule[-1.5ex]{0pt}{3ex} \multirow{2}{*}{} & \multicolumn{2}{c|}{\small \textbf{c-Fos$^+$ cell segmentation}} & \multicolumn{4}{c}{\small \textbf{microglia segmentation}} \\
\rule[-1.5ex]{0pt}{3ex} {Teams} & {\small $Dice(\%) \uparrow$} & {\small $\beta_0$ error $\downarrow$} & {\small $Dice(\%) \uparrow$} & {\small $clDice (\%) \uparrow$}& {\small $\beta_0$ error $\downarrow$} & {\small $\beta_1$ error $\downarrow$}\\
\hline
\rule[-1.5ex]{0pt}{3ex} {\footnotesize Zoomlin}& {\footnotesize{$\boldsymbol{75.00\pm8.47}$}} & {\footnotesize{$38.48\pm49.35$}} & {\footnotesize{$\boldsymbol{71.50\pm15.02}$}} & {\footnotesize{$74.10\pm14.55$}} & {\footnotesize{$34.09\pm18.57$}} & {\footnotesize{$\boldsymbol{0.2471\pm0.5293}$}} \\
\rule[-1.5ex]{0pt}{3ex} {\footnotesize BioAI}& {\footnotesize{$69.27\pm 9.62$}} & {\footnotesize{$97.78\pm82.60$}} & {\footnotesize{$66.17\pm17.52$}} & {\footnotesize{$75.44\pm18.31$}} & {\footnotesize{$\boldsymbol{11.73\pm7.55}$}} & {\footnotesize{$0.6353\pm1.3534$}} \\
\rule[-1.5ex]{0pt}{3ex} {\footnotesize XunDJ}& {\footnotesize{$63.30\pm 15.44$}} & {\footnotesize{$70.39\pm61.76$}} & {\footnotesize{ $69.15\pm16.09$}} & {\footnotesize{$\boldsymbol{76.73\pm17.19}$}}& {\footnotesize{$14.92\pm12.18$}} & {\footnotesize{$0.5529\pm0.9882$}} \\
\rule[-1.5ex]{0pt}{3ex} {\footnotesize Tonyxu}& {\footnotesize{$68.60\pm 11.61$}} & {\footnotesize{$\boldsymbol{34.04\pm26.43}$}} & {\footnotesize{$65.88\pm17.49$}} & {\footnotesize{$71.48\pm18.08$}} & {\footnotesize{$19.54\pm12.04$}} & {\footnotesize{$0.6706\pm1.6116$}} \\
\rule[-1.5ex]{0pt}{3ex} {\footnotesize Wu}& {\footnotesize{$35.75\pm 11.12$}} & {\footnotesize{$174.1\pm194.7$}} & {\footnotesize{ $47.78\pm8.46$}} & {\footnotesize{$43.10\pm12.46$}}& {\footnotesize{$97.51\pm50.46$}} & {\footnotesize{$0.4000\pm0.5358$}} \\
\hline
\end{tabular}
\end{table}

\section{Results}
In this section, we report the results of participant methods during both the preliminary and final test phases. We conduct both quantitative and qualitative assessments of the segmentation results. Through the comparison, we aim to provide a robust evaluation of the participant methods, offering valuable insights into their performance. Finally, we present results from some participants, highlighting the performance differences with and without the self-supervised learning stage, to demonstrate the value and impact of self-supervised learning.

\subsection{Preliminary Test Phase}
\begin{figure}[h!]
    \centering
    \includegraphics[width=1.0\textwidth]{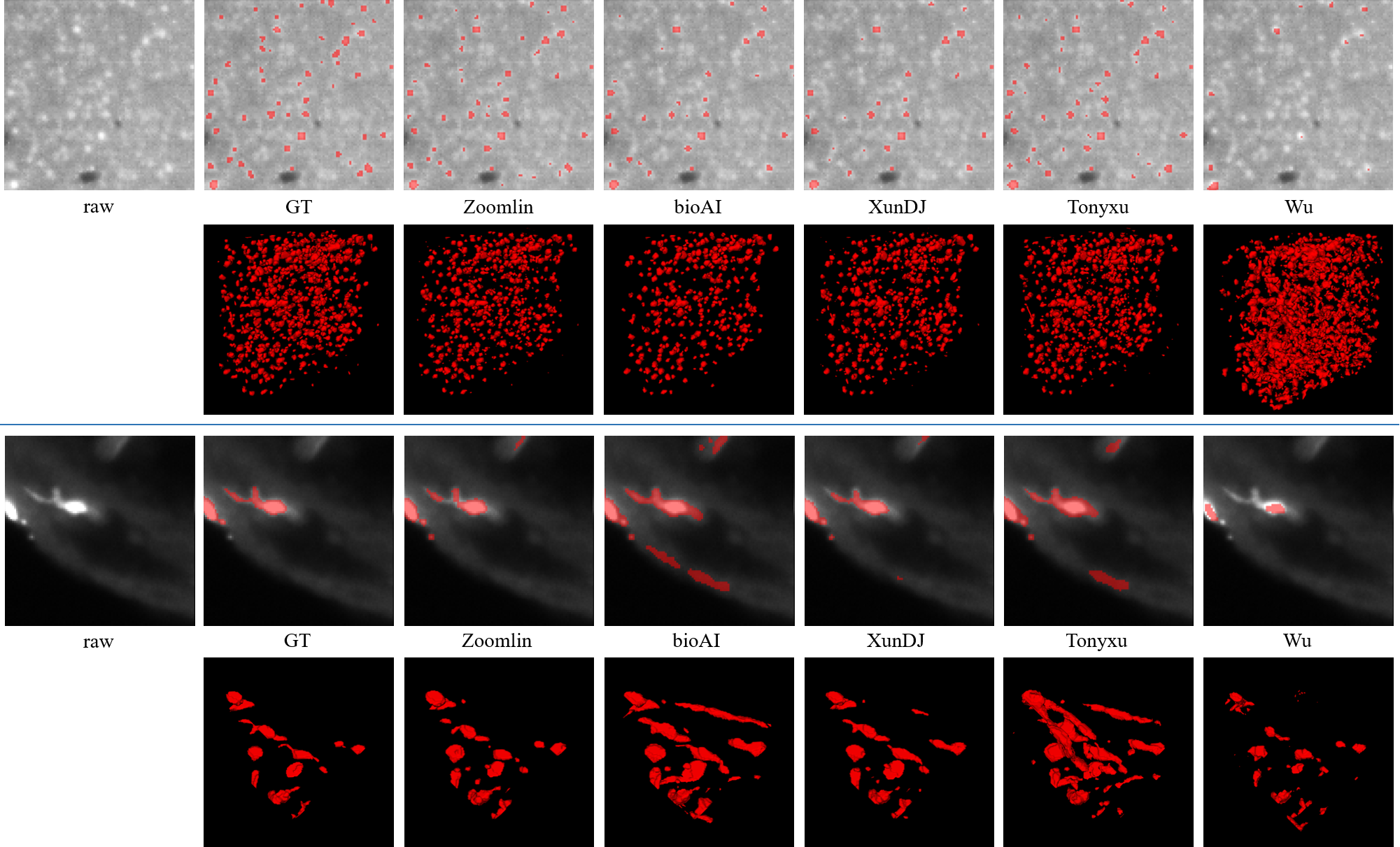} 
    \caption{Qualitative results for c-Fos$^+$ cell segmentation (upper panel) and microglia segmentation (lower panel) during the preliminary test phase. In each panel, the first row displays the segmentation results in a 2D slice view (segmentation highlighted in red overlaid on the raw images), while the second row provides a 3D view of the segmentation results.}
    \label{fig:preliminary}
\end{figure}

In the preliminary test phase, five teams participated. Table \ref{tab:pre_results} presents the quantitative assessment of the segmentation results achieved by these five teams. Figure \ref{fig:preliminary} showcases examples of c-Fos$^+$ cell segmentation and microglia segmentation from each team. All results were obtained from the best submission by each team prior to the preliminary phase deadline.

The Zoomlin team achieved the highest $Dice$ scores for both c-Fos$^+$ cell segmentation and microglia segmentation, indicating that their results were the most closely aligned with the ground truth. This is further supported by the examples presented in Figure \ref{fig:preliminary}, where their segmentation results are visually consistent with the ground truth in most parts. Both Zoomlin and Tonyxu teams demonstrated strong performance in terms of low $\beta_0$ errors for c-Fos$^+$ cell segmentation. As illustrated in Figure \ref{fig:preliminary}, Zoomlin and Tonyxu teams  effectively detected the majority of c-Fos$^+$ cells, showing a high level of sensitivity. In contrast, the methods of bioAI and XunDJ teams missed a significant number of cells. On the other hand, the Wu team’s method caused misaligned segmentations when compared to the ground truth. Overall, Zoomlin and Tonyxu teams performed better than other teams in terms of c-Fos$^+$ cell segmentation. The differences in performance across the teams highlight varying levels of 
effectiveness in detecting and segmenting the c-Fos$^+$ cell structures.

For microglia segmentation, the Zoomlin, BioAI and XunDJ teams achieved high $clDice$ scores, highlighting their ability in preserving the tubular structures of microglia branches. The Zoomlin team, in particular, attained the lowest $\beta_1$ errors, further confirming the effectiveness of their method in accurately capturing the shape of microglia. However, the Zoomlin team exhibited higher $\beta_0$ errors compared to the BioAI, XunDJ, and Tonyxu teams. This suggests that segmentations from the Zoomlin team exhibit discontinuities, as also evident in the segmentation results presented in Figure \ref{fig:preliminary}.

\subsection{Final Test Phase}
\begin{figure}[t!]
    \centering
    \includegraphics[width=1.0\textwidth]{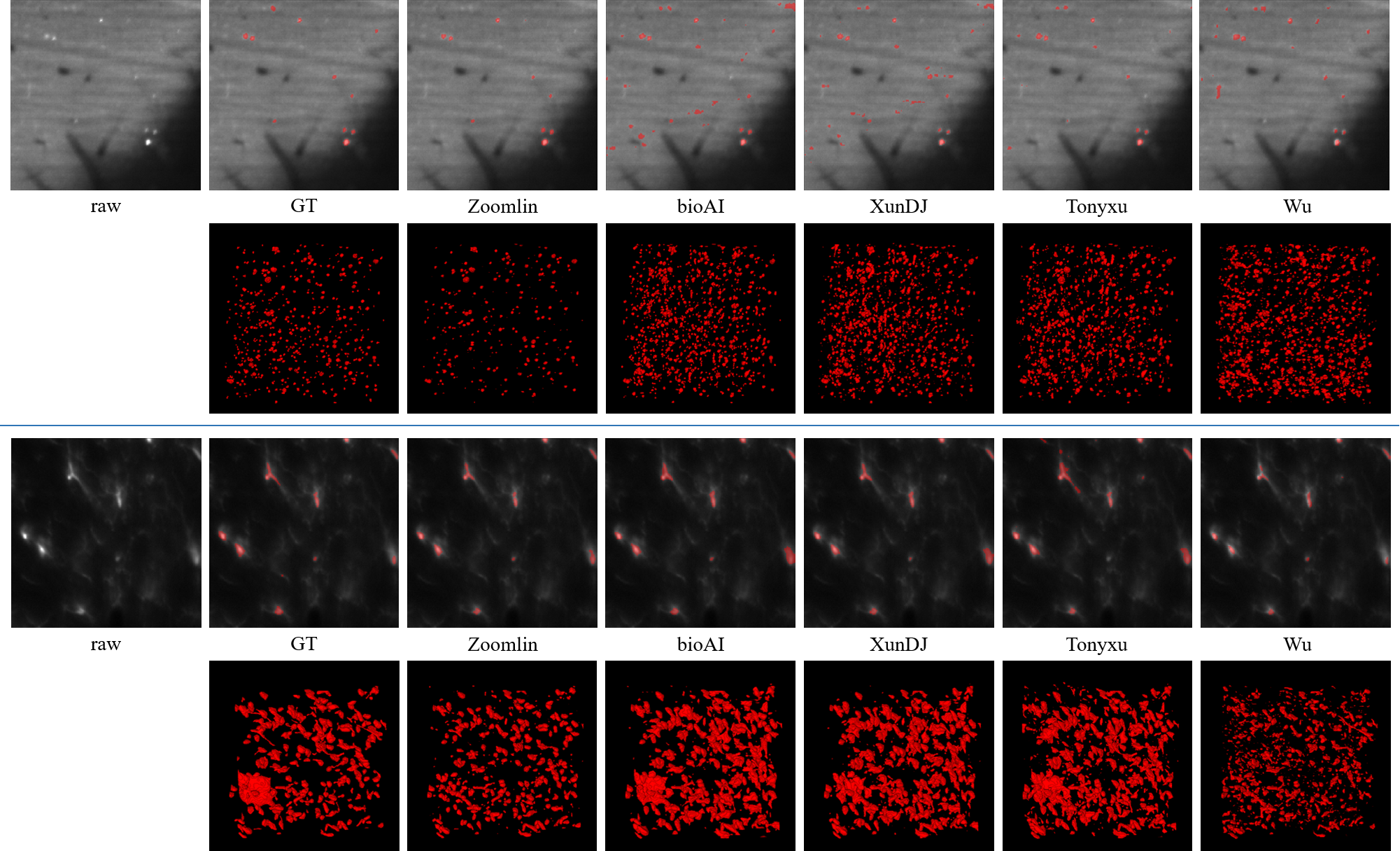} 
    \caption{Qualitative results for c-Fos$^+$ cell segmentation (upper panel) and microglia segmentation (lower panel) during the final test phase. In each panel, the first row presents the segmentation displays in a 2D slice view (segmentation highlighted in red overlaid on the raw images), while the second row provides a 3D view of the segmentation results.}
    \label{fig:final}
\end{figure}

Five teams participated in the final test phase. Table \ref{tab:final_results} presents the quantitative evaluation of segmentation results obtained by these teams. Figure \ref{fig:final} provides examples of c-Fos$^+$ cell segmentation and microglia segmentation performed by each team.

\begin{table}[t]
\caption{Quantitative results on the final test sets achieved by participants.} 
\label{tab:final_results}
\centering
\begin{tabular}{c c c | c c c c} 
\hline
\rule[-1.5ex]{0pt}{3ex} \multirow{2}{*}{} & \multicolumn{2}{c|}{\small \textbf{c-Fos$^+$ cell segmentation}} & \multicolumn{4}{c}{\small \textbf{microglia segmentation}} \\
\rule[-1.5ex]{0pt}{3ex} {Teams} & {\small $Dice(\%) \uparrow$} & {\small $\beta_0$ error $\downarrow$} & {\small $Dice(\%) \uparrow$} & {\small $clDice (\%) \uparrow$}& {\small $\beta_0$ error $\downarrow$} & {\small $\beta_1$ error $\downarrow$}\\
\hline
\rule[-1.5ex]{0pt}{3ex} {\footnotesize Zoomlin}& {\footnotesize{$\boldsymbol{65.37\pm13.38}$}} & {\footnotesize{$\boldsymbol{157.9\pm169.9}$}} & {\footnotesize{$\boldsymbol{65.17\pm9.11}$}}  & {\footnotesize{$\boldsymbol{73.77\pm13.17}$}} & {\footnotesize{$\boldsymbol{125.5\pm47.5}$}}  & {\footnotesize{$\boldsymbol{1.000\pm0.8660}$}} \\
\rule[-1.5ex]{0pt}{3ex} {\footnotesize BioAI}&  {\footnotesize{$56.21\pm4.70$}} & {\footnotesize{$238.8\pm98.5$}} & {\footnotesize{$51.24\pm19.95$}} & {\footnotesize{$59.72\pm21.43$}}& {\footnotesize{$145.3\pm58.7$}} & {\footnotesize{$5.125\pm2.848$}} \\
\rule[-1.5ex]{0pt}{3ex} {\footnotesize XunDJ}& {\footnotesize{$50.67\pm15.10$}} & {\footnotesize{$549.5\pm373.8$}} & {\footnotesize{$54.04\pm17.32$}} & {\footnotesize{$61.40\pm20.15$}}& {\footnotesize{$184.4\pm99.5$}} & {\footnotesize{$6.250\pm3.152$}} \\
\rule[-1.5ex]{0pt}{3ex} {\footnotesize Tonyxu}& {\footnotesize{$61.55\pm4.17$}} & {\footnotesize{$204.6\pm159.1$}} & {\footnotesize{$48.69\pm20.50$}} & {\footnotesize{$55.73\pm20.56$}}& {\footnotesize{$226.0\pm136.1$}} & {\footnotesize{$5.375\pm3.160$}} \\
\rule[-1.5ex]{0pt}{3ex} {\footnotesize Wu}& {\footnotesize{$37.22\pm4.69$}} & {\footnotesize{$743.5\pm407.9$}} & {\footnotesize{$48.11\pm6.41$}} & {\footnotesize{$39.80\pm16.29$}}& {\footnotesize{$692.3\pm471.9$}} & {\footnotesize{$2.125\pm1.965$}} \\
\hline
\end{tabular}
\end{table}

The Zoomlin team excelled in the c-Fos$^+$ cell segmentation task, achieving the best scores in terms of both $Dice$ and $\beta_0$ errors. Their results indicate superior accuracy in detecting and segmenting c-Fos$^+$ cells. The Tonyxu team also attained a higher $Dice$ score and lower $\beta_0$ errors compared to the other teams, suggesting their method's superior performance in c-Fos$^+$ cell segmentation. As shown in Figure \ref{fig:final}, both the Zoomlin and Tonyxu teams demonstrated high accuracy in detecting the majority of c-Fos$^+$ cells, whereas the BioAI, XunDJ, and Wu teams produced a considerable number of false positives. This performance aligns with the trend observed in the preliminary test phase, where the Zoomlin and Tonyxu teams consistently outperformed the others in c-Fos$^+$ cell segmentation task.

In the microglia segmentation task, team Zoomlin achieved the best scores across all metrics. Meanwhile, in the preliminary test phase, the Zoomlin team outperformed other groups in terms of two metrics. Given that microglia represent a new biological structure not included in the training set, these results underscore the exceptional generalizability of Zoomlin's model, distinguishing it from those of the other teams. This ability to accurately segment previously unseen structures highlights the robustness and adaptability of their approach.

\subsection{Impact of Self-supervised Learning}
The Zoomlin team, ranked first, demonstrated superior performance in both c-Fos$^+$ cell segmentation and microglia segmentation during both the preliminary and final test phases. Notably, their successful segmentation of the unseen structure, microglia, further demonstrated the generalizability of their model. They also provided a comparison of their model's performance with and without the self-supervised learning stage for the preliminary test phase. However, since each participant was allowed only one submission for the final test phase, they were unable to provide the same comparison for the final test set.

\setlength{\tabcolsep}{6pt}
\begin{table}[t]
\caption{Quantitative results of the Zoomlin team's segmentation model, with and without self-supervised learning (SSL), on the preliminary test set. } 
\label{tab:ssl}  
\centering
\begin{tabular}{c c c | c c c c} 
\hline
\rule[-1.5ex]{0pt}{3ex} \multirow{2}{*}{} & \multicolumn{2}{c|}{\small \textbf{c-Fos$^+$ cell segmentation}} & \multicolumn{4}{c}{\small \textbf{microglia segmentation}} \\
\rule[-1.5ex]{0pt}{3ex} & {\small $Dice(\%) \uparrow$} & {\small $\beta_0$ error $\downarrow$} & {\small $Dice(\%) \uparrow$} & {\small $clDice (\%) \uparrow$}& {\small $\beta_0$ error $\downarrow$} & {\small $\beta_1$ error $\downarrow$}\\
\hline
\rule[-1.5ex]{0pt}{3ex} {\footnotesize w/ SSL}& {\footnotesize {$75.00\pm8.47$}} & {\footnotesize {$38.48\pm49.35$}} & {\footnotesize {$71.50\pm15.02$}} & {\footnotesize {$74.10\pm14.55$}} & {\footnotesize {$34.09\pm18.57$}} & {\footnotesize {$0.2471\pm0.5293$}} \\
\rule[-1.5ex]{0pt}{3ex} {\footnotesize wo/ SSL }& {\footnotesize {$76.43\pm 8.07$}} & {\footnotesize {$38.30\pm43.47$}} & {\footnotesize {$69.87\pm14.92$}} & {\footnotesize {$68.08\pm15.35$}} & {\footnotesize {$59.24\pm31.30$}} & {\footnotesize {$0.3412\pm0.6049$}} \\
\hline
\end{tabular}
\end{table}

Table \ref{tab:ssl} displays the quantitative results of the Zoomlin team's segmentation model in the preliminary test phase, both with and without the self-supervised learning stage. It is evident that the model showed significant improvement across all metrics for microglia segmentation after employing self-supervised learning. This performance boost in microglia segmentation provides evidence for the effectiveness of self-supervised learning in enhancing the model's robustness and generalization capabilities. However, the segmentation model's performance decreased slightly after employing self-supervised learning. This suggests that while the same self-supervised learning strategy can provide significant benefits for certain tasks, its impact may be limited for others. For the Zoomlin team, their self-supervised learning approach enhanced performance for tree-like structure segmentation, but led to a reduction in accuracy for spot-like structure segmentation. 

Based on the results from Zoomlin's team and considering the significant differences in topology and segmentation focus between spot-like and tree-like structures, we propose that effective self-supervised learning strategies should be tailored to each type of structure. These strategies should be designed according to the specific structural characteristics of the structures being segmented. 

\section{Conclusion}
In this paper, we presented the Self-Supervised Learning for 3D Light-Sheet Microscopy Image Segmentation (SELMA3D) challenge, which was  held at MICCAI 2024. In this challenge, we provided a substantial collection of 3D light-sheet microscopy (LSM) images, along with a subset of annotated patches. The biological structures in 3D LSM images were categorized into two types: spot-like structures and tree-like structures. We evaluated the participants' methods on unseen biological structures which fell into the two types. 
The top-performing model achieved over $70\%$ Dice score for both spot-like and tree-like structure segmentation in the preliminary test phase, and over $65\%$ Dice score for both tasks in the final test phase. Compared to fully supervised baselines, the participants methods leveraging self-supervised learning achieved better performance. 
This clear observation reflects a succesful outcome of the challenge, which is the development of better and more generalizable models through self supervised learning. 

However, the experimental results also revealed that the benefits of self-supervised learning for one type of structure do not always translate into consistent performance improvements for other types. This indicates that the effectiveness of self-supervised learning is dependent on the structural characteristics of the objects being segmented. Consequently, we conclude that it can be beneficial to design self-supervised learning strategies that are specifically tailored to different types of structures, whether they are spot-like or tree-like. In light of these findings, our future work will focus on developing self-supervised learning strategies that are specific and shape aware to optimally reconstruct different structural features. 

\section{Data and Code Availability}
For information on accessing the data, evaluation code, and submmited Docker containers from participants in this challenge, please visit the challenge's homepage \href{https://selma3d.grand-challenge.org/}{https://selma3d.grand-challenge.org/}. 

\section{Acknowledgments}
The challenge is supported by the Institute for Tissue Engineering and Regenerative Medicine (iTERM) at Helmholtz Zentrum München. We thank Harsharan S. Bhatia, Doris Kaltenecker, Mihail Todorov and Moritz Negwer for the assistance in data collection. We also thank Alex Berger and Martin Menten for their discussions and suggestions on the proposal, as well as Laurin Lux for testing the evaluation method. The authors would like to thank the MICCAI challenge society and the support of Grand Challenge.

\section{References}


\bibliography{report}   
\bibliographystyle{spiejour}   


\end{spacing}
\end{document}